# Theoretical and Experimental Investigation into the Branched Flow Phenomenon of Light

*Han Lin[1] [\*]*

[1]*School of Atmospheric Sciences, Nanjing University (Xianlin Campus)*
*Nanjing, 210023, China*

*Xiaoyue Ma[2, \*, #]*

[2]*School of Environment, Nanjing University (Xianlin Campus)*
*Nanjing, 210023, China*

*Kefan Wang[3]*

[3]*School of Physics, Nanjing University (Xianlin Campus)*
*Nanjing, 210008, China*

**Abstract** Branched flow can be observed when a laser beam is coupled into a soap film. This research theoretically explored the phenomenon through analogy between light wave and particles in form of Hamilton-Jacobian equation, further discussed the shape and statistics of the branched flow with Fokker-Plank equation and experimentally adopted the methods of interference imaging, computer image processing and so on to obtain the experiment data that could prove previously deduced scaling law concerning the statistical properties of the branched flow.

## § 1. Introduction

When waves or rays travel through weak disordered media with proper correlation length larger than their typical length scales (which, for waves, are their wavelengths), between the starting point and the totally chaotic stage there exists an intermediate regime where the waves or rays are diffusive 'only in an ensemble sense'.[1] In this intermediate regime, the weakly refracted waves or rays form focal 'cusps' and rogue waves through interference. This phenomenon has been observed in systems involving diverse length scales, including electron gas [2], tsunamis [3], pulsar stars and even the large-scale structure of the universe. This topic has long been the research focus of Professor E. J. Heller and his team, whose provocative works can be viewed altogether on the website in Ref. [4].

Nevertheless, due to the heterochromatic nature of most light source researchers use in labs and the lack of a proper wave guide, the branching phenomenon had not been observed until in 2020 a research team in Israel accidentally discovered its existence when conducting irrelevant experiments.[5] The team then designed and carried out specific experiments to examine its statistical properties.

Later, in 2021 IYPT (International Young Physicists' Tournament) problems, branched flow of light appeared as question NO.9 with the name 'Light Whiskers'.[6] The problem

---

[\*] These authors contributed equally to this work.
[#] 201830014@smail.nju.edu.cn

reads that a laser beam entering a thin soap film at a small angle would exhibit the form of a rapidly changing thin branching light path, and the participants were asked to explain and investigate the phenomenon. The IYPT problems were at the same time adopted for CUPT (China Undergraduate Physics Tournament), in which the Nanjing University team took part.

During the research process, Han Lin from School of Atmospheric Sciences, Nanjing University greatly enriched the deduction presented by A. Patsyk et al. in their Supplementary Material[5] with an analog between wave and particle which enabled analysis from a Hamilton-Jacobian viewpoint, and in the meantime helped specified the physical quantities that could later be measured by experiment, namely the potential field of the soap film in its effective refraction index, mean refraction index and the wave vector of the incident laser beam; the correlation length of the random potential; the characteristic length of the branched flow etc.

Xiaoyue Ma, from School of Environment, Nanjing University, also the general leader of the research team, then designed and conducted certain experiments, gathered, analyzed and visualized data with technical help from Kefan Wang of School of Physics, Nanjing University. The research process and conclusion are as listed below.

## § 2. Theoretical Background

In this part, the branching phenomenon of light is analyzed both from a general branched-flow perspective and in a specific light-wave context.

### § 2.1 the Kick-Drift Cycle model and the Universal Scaling Law

According to E. J. Heller et al [1], all branched flow could be viewed as rays being refracted by independent thin lenses/phase screens along their traveling path. During this process there is a constant cycle of 'kick-drift': the ray would drift with its current momentum for a while, then receive a random impulsive momentum kick normal to the mean flow, then drift again, and then kick again.

This could be described in the language of mathematics as a point-to-point area preserving self-mapping phase plane, under the relation:

$$p_{n+1} = p_n - \frac{dV_n(x)}{dx}|_{x=x_n}, \qquad (kick\ step)$$

$$q_{n+1} = q_n + p_{n+1}, \qquad (drift\ step)$$

where $p$ denotes the momentum of the ray, $x$ the position, $V$ the weak-ordered potential and $n$ the discretized time.

Apart from this model, another property universally observed in branched flow is that the correlation length $l_c$ of the wave guide (i.e., the potential field in the kick-drift cycle model) has the following relation to the characteristic length $L$ (i.e., the distance from the starting point to the first 'cusp') of the waves or rays:

$$L = l_c \left(\frac{E}{\varepsilon}\right)^{\frac{2}{3}}. \qquad (2.1.1)$$

In the above equation $E$ denotes the typical energy of the system and $\varepsilon$ the random variation of the potential with relation $\varepsilon \ll E$.

### § 2.2 Working out the Scaling law

#### § 2.2.1 Coupling of Laser and Soap Film

In the experiment, the inserted laser beam can be seen as a monochromatic wave packet with frequency $\omega = ck_0$. Since the typical length the film stretches over is much larger than the wavelength of the laser, for now, the soap film can be modelled as a dielectric slab guide. The refractive index inside the slab is n and the index of the surrounding air is 1.

We take an approach different from the one used in Ref. [5], by considering one Fourier component of the initial wave packet bouncing between the slab surfaces, see Fig.1. In the experiments, the incident angle $\theta_i$ is a small angle, allowing the appearance of total reflection.

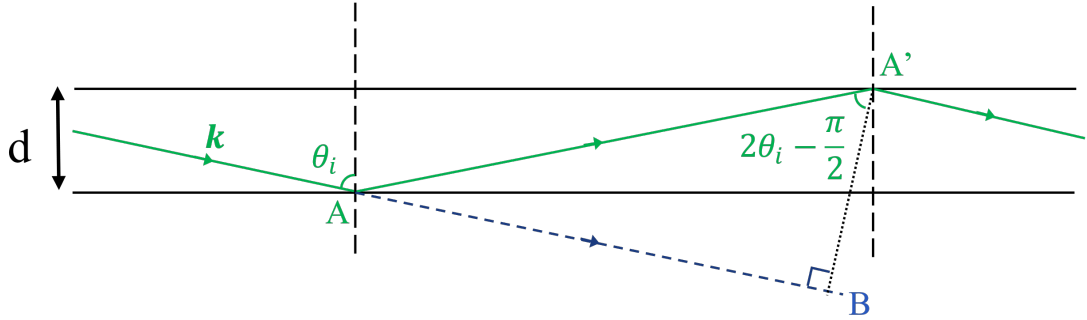

Fig. 1 Plane wave propagating in the slab waveguide

For the plane wave to travel steadily, the accumulation of phase during the path from A to just pass B (marked in green, with different shades denoting the effect of reflections) must be identical to the imaginary path from A to C (marked in blue), up to an integer multiple of $2\pi$. So, the phase difference between two paths is given by:

$$\Delta\phi = \frac{k_0 nd}{\cos\theta_i} - \frac{k_0 nd}{\cos\theta_i}\cos(\pi - 2\theta_i) - 2\cdot 2\theta = 2p\pi, p = 0,1,2,\ldots \quad (2.2.1.1)$$

For simplicity, we consider the case of TE-polarized mode first. The phase change $2\theta$ for each of the reflections is given by the Fresnel formulas:

$$E_r = \frac{n\cos\theta_i - \cos\theta_t}{n\cos\theta_i + \cos\theta_t}E_i = \frac{1-i\alpha}{1+i\alpha}E_i = e^{-2i\theta}E_i, \quad (2.2.1.2)$$

where

$$\alpha = \frac{\sqrt{n^2\sin^2\theta_i - 1}}{n\cos\theta_i}, \theta = \arctan\alpha \quad (2.2.1.3)$$

Simple calculation yields

$$k_0 nd\sqrt{n^2 - n^2\sin^2\theta_i} + 2\arctan\sqrt{\frac{n^2 - n^2\sin^2\theta_i}{n^2\sin^2\theta_i - 1}} = (p+1)\pi \quad (2.2.1.4)$$

with $p = 0, 1, 2 \ldots$

The relationship imposes strong constraint on the transverse components of the wave vector. Specifically speaking, they must satisfy the relationship

$$\frac{|\mathbf{k} - (\mathbf{k}\cdot\hat{\mathbf{n}})\hat{\mathbf{n}}|}{k_0} = n_{eff} \quad (2.2.1.5)$$

where $n_{eff}$ is the solution(s) to the equation for $n\sin\theta_i$ and $\hat{\mathbf{n}}$ is the unit vector perpendicular to the slab plane. The wave equation for the scalar electric field must be

$$\nabla_T^2 \Psi + k_0^2 n_{eff}^2 \Psi = \left(\frac{\partial^2}{\partial x^2} + \frac{\partial^2}{\partial z^2}\right)\Psi + k_0^2 n_{eff}^2 \Psi = 0 \quad (2.2.1.6)$$

since the same equation holds for every Fourier component of the scalar field.

The derivation above can be applied to the TM-polarized wave with minor modification. The result is:

$$k_0 nd \sqrt{n^2 - n_{eff}^2} + 2\,arctan\left(\frac{1}{n^2}\sqrt{\frac{n^2 - n_{eff}^2}{n_{eff}^2 - 1}}\right) = (p+1)\pi, p = 0, 1, 2 \dots \quad (2.2.1.7)$$

**§ 2.2.2 Shape of the Branches**

In real world, the thickness of the soap film is slowly varying with space and time. We assume that the relationship derived above still holds as a result of adiabatic approximation. The exact distribution of n_{eff} can be measured through experiments and is capable to support numerical simulations. However, we wish to understand some key features of the outcome in advance.

Hence, we shall view the effective refraction index as a random function of space coordinates, and study how light behaves when propagating through such a medium.

To study the shape of the branches, we derive the ray-limit result from the field equation.

First, rewrite the field in terms of its amplitude and phase explicitly, and plug it into the field equation:

$$\psi = ae^{iS} \Rightarrow \nabla\psi = (\nabla a + ia\nabla S)e^{iS}$$

$$\Rightarrow \nabla^2\psi = [\nabla^2 a + i\nabla\cdot(a\nabla S) + i\nabla a\cdot\nabla S - a(\nabla S)^2]e^{iS} = -k_0^2 n_{eff}^2 a e^{iS} \quad (2.2.2.1)$$

Take the imaginary and real part, the equation becomes:

$$\Rightarrow (\nabla S)^2 = k_0^2 n_{eff}^2 - \frac{\nabla^2 a}{a}, \qquad (real\ part)$$

$$a\nabla\cdot(a\nabla S) + \nabla a\cdot a\nabla S = \nabla\cdot(a^2\nabla S) = 0. \qquad (imaginary\ part)$$

In the ray limit, the amplitude acts as a passive scalar field and we arrive at the Hamilton-Jacobi equation:

$$(\nabla S)^2 = k_0^2 n_{eff}^2, \nabla\cdot(a^2\nabla S) = 0 \quad (2.2.2.2)$$

Assuming the ray is initially oriented at the z direction and doesn't defect much during propagation due to the weak background potential, the phase can be split into two parts.

$$S = k_0 \bar{n} z + S_1(x, z) \quad (2.2.2.3)$$

One with linear z dependence which merely reflects the phase accumulation when propagating along the original direction. The other one, which we denote by S_1, describes the shape of the ray and is what we are truly interested in. Assuming S_1 varies slowly with z and its second-order derivatives can be neglected:

$$(\nabla_T S)^2 = \left[k_0^2\bar{n}^2 + \frac{2k_0\bar{n}(\partial S_1)}{\partial z} + \left(\frac{\partial S_1}{\partial z}\right)^2\right] + \left(\frac{\partial S_1}{\partial x}\right)^2 \approx k_0^2\bar{n}^2 + \frac{2k_0\bar{n}(\partial S_1)}{\partial z} + \left(\frac{\partial S_1}{\partial x}\right)^2$$

$$\Rightarrow \frac{\partial S_1}{\partial z} + \frac{1}{2k_0\bar{n}}\left(\frac{\partial S_1}{\partial x}\right)^2 + \frac{k_0}{2\bar{n}}(\bar{n}^2 - n_{eff}^2) = 0 \quad (2.2.2.4)$$

This equation looks exactly like the Hamilton-Jacobi equation describing the motion of a single particle: $\frac{\partial S_1}{\partial t} + \frac{1}{2m}\left(\frac{\partial S_1}{\partial x}\right)^2 + V(x,t) = 0$, with simple replacements:

$$z = y, m = k_0\bar{n}, V(x,z) = \frac{k_0}{2\bar{n}}[\bar{n}^2 - n_{eff}^2(x,z)] \quad (2.2.2.5)$$

As mentioned before, we treat the effective

$$\begin{cases} \dot{x} = \dfrac{p}{m} \\ \dot{p} = -\dfrac{\partial V}{\partial x} \end{cases} \Rightarrow \frac{\partial \rho}{\partial t} + \frac{\partial (\dot{x}\rho)}{\partial x} + \frac{\partial (\dot{p}\rho)}{\partial p} = 0 \qquad (2.2.2.6)$$

$$\frac{\partial \rho}{\partial t} + \frac{p}{m}\frac{\partial \rho}{\partial x} = \frac{\partial V}{\partial x}\frac{\partial \rho}{\partial p} \Rightarrow \frac{\partial}{\partial t}\left[e^{t\hat{L}}\rho\right] = e^{t\hat{L}}\frac{\partial V}{\partial x}\frac{\partial \rho}{\partial p}, \hat{L} \equiv \frac{p}{m}\frac{\partial}{\partial x} \qquad (2.2.2.7)$$

$$\rho = e^{-t\hat{L}}\rho_0 + \int dt' e^{-(t-t')\hat{L}} \frac{\partial V}{\partial x}\frac{\partial \rho}{\partial p} \qquad (2.2.2.8)$$

Where $\rho_0$ is the initial distribution function. Plug this formal solution back in the original Liouville's equation:

$$\frac{\partial \rho}{\partial t} + \frac{p}{m}\frac{\partial \rho}{\partial x} = \frac{\partial V}{\partial x}\frac{\partial}{\partial p}\left[e^{-t\hat{L}}\rho_0 + \int dt' e^{-(t-t')\hat{L}}\frac{\partial V}{\partial x}\frac{\partial \rho}{\partial p}\right] \qquad (2.2.2.9)$$

To

$$\frac{\partial \langle\rho\rangle}{\partial t} + \frac{p}{m}\frac{\partial \langle\rho\rangle}{\partial x} = \left\langle\frac{\partial V}{\partial x}\frac{\partial}{\partial p}\left(e^{-t\hat{L}}\rho_0\right)\right\rangle + \left\langle\frac{\partial V}{\partial x}\frac{\partial}{\partial p}\left[\int dt' e^{-(t-t')\hat{L}}\frac{\partial V}{\partial x}\frac{\partial \rho}{\partial p}\right]\right\rangle \qquad (2.2.2.10)$$

Approximate the effect of the random potential in the following manner, such that the particles undergo a Markov process

$$\left\langle\frac{\partial V}{\partial x}\right\rangle = 0, \left\langle\frac{\partial V(x,t)}{\partial x}\frac{\partial V(x',t')}{\partial x'}\right\rangle \approx \left\langle\frac{\partial V(x,t)}{\partial x}\frac{\partial V(x',t')}{\partial x'}\right\rangle\bigg|_{x=x'=0} \approx \sigma^2 \delta(t-t') \qquad (2.2.2.11)$$

Where the corresponding coefficient sigma can be calculated in the following way:

$$\sigma^2 = \int dt' \sigma^2 \delta(t-t') \approx \int dt' \left\langle\frac{\partial V(x,t)}{\partial x}\frac{\partial V(x',t')}{\partial x'}\right\rangle\bigg|_{x=x'=0} \qquad (2.2.2.12)$$

with

$$\left\langle\frac{\partial V(x,t)}{\partial x}\frac{\partial V(x',t')}{\partial x'}\right\rangle\bigg|_{x=x'=0} = \frac{\partial^2 \langle V(x,t)V(x',t')\rangle}{\partial x \partial x'}\bigg|_{x=x'=0}$$

$$= -\frac{\partial^2}{\partial x^2} c(x-x', t-t')\bigg|_{x=x'=0} \qquad (2.2.2.13)$$

We have

$$\sigma^2 = -\int dt \frac{\partial^2 c(x,t)}{\partial x^2}\bigg|_{x=0} \sim \frac{\sigma_V^2}{l_c}, \sigma_V^2 = \langle V^2(x)\rangle = c(0) \qquad (2.2.2.14)$$

$$\frac{\partial \langle\rho\rangle}{\partial t} = -\frac{p}{m}\frac{\partial \langle\rho\rangle}{\partial x} + \sigma^2 \frac{\partial^2 \langle\rho\rangle}{\partial p^2} \qquad (2.2.2.15)$$

This is the desired Fokker-Planck equation. Now we're ready to work out the second-moments:

$$\frac{d}{dt}\langle x^2\rangle = \int dxdp \frac{\partial \langle\rho\rangle}{\partial t} x^2 = \int dxdp \left(-\frac{p}{m}\frac{\partial \langle\rho\rangle}{\partial x} + \sigma^2 \frac{\partial^2 \langle\rho\rangle}{\partial p^2}\right) x^2 = \int dxdp \frac{2xp}{m}\langle\rho\rangle$$

$$= \frac{2}{m}\overline{xp} \qquad (2.2.2.16)$$

$$\frac{d}{dt}(\overline{xp}) = \int dxdp\left(-\frac{p}{m}\frac{\partial\langle\rho\rangle}{\partial x} + \sigma^2\frac{\partial^2\langle\rho\rangle}{\partial p^2}\right)xp = \frac{1}{m}\overline{p^2} \qquad (2.2.2.17)$$

$$\frac{d}{dt}(\overline{p^2}) = \int dxdp\left(-\frac{p}{m}\frac{\partial\langle\rho\rangle}{\partial x} + \sigma^2\frac{\partial^2\langle\rho\rangle}{\partial p^2}\right)p^2 = 2\sigma^2 \qquad (2.2.2.18)$$

Combining all the results above and solve the final differential equation, we have

$$\overline{x^2} = \frac{2\sigma^2}{3m^2}t^3 = \frac{2\sigma^2}{3m^2}z^3$$

$$l_c^2 \sim \frac{2\sigma_V^2}{3m^2 l_c}z^3 \sim \frac{\sigma_V^2}{m^2 l_c}z^3 \Rightarrow z \sim l_c\left(\frac{m^2}{\sigma_V^2}\right)^{\frac{1}{3}} = l_c\left(\frac{m}{\sigma_V}\right)^{\frac{2}{3}} \qquad (2.2.2.19)$$

$$E = m = k_0\,n, \varepsilon = \sigma_V \qquad (2.2.2.20)$$

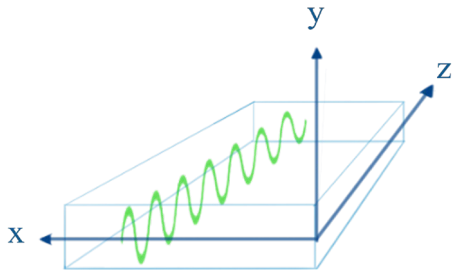

Fig. 2 Coordinates

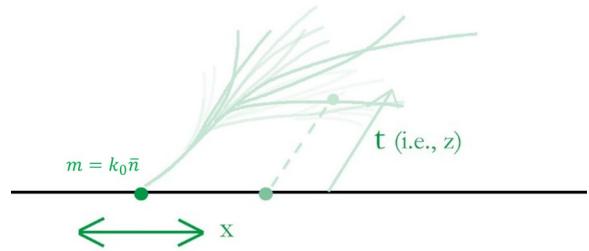

Fig. 3 Analogy from wave to particle

## § 3. Experimental Methods & Setup

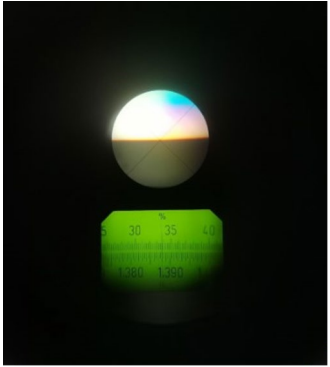

Fig. 4 The Abel refractometer

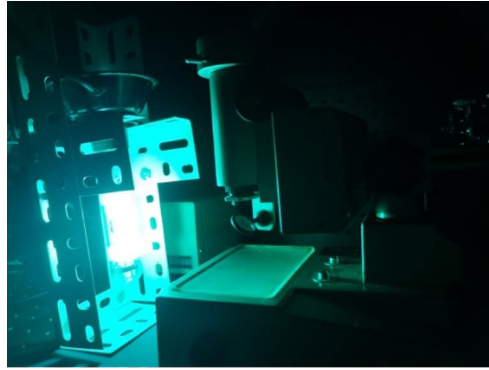

Fig. 5 Equipment setup for the interference experiment

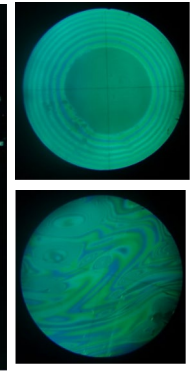

Fig. 6 Newton's Ring, and the interferometric pattern of a soap film

    Firstly, pre-experiments were conducted to obtain certain proportion of chemicals to produce long-lasting, thin soap films. The thickness was roughly measured by observing the interference pattern and color in the soap film against white room light by naked eyes, and the endurance of certain liquid was measured by timing the soap film from its stabilization to its breaking and take the average value of multiple test results. In the end, the chosen recipe was: $50mL$ water to provide the structure, $10mL$ glycerin to prolong its life, $30mL$ bubble water

to make thin the soap film, $10mL$ detergent also to prolong its life, and $30g$ soft sugar to add to the viscosity.

Secondly, the inner refractive index of the basic liquid was measure with an Abel refractometer with environment temperature at 25~27℃.

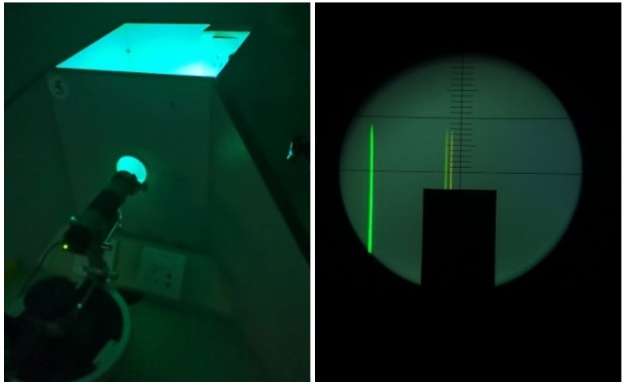

Fig. 7 Spectrometer, and the spectrum of Hg lamp

Thirdly, the interference images of the soap film were obtained using a Hg lamp and a reading microscope to calculate the thickness distribution of the soap film. The plastic hoop bracing the soap film was self-designed and 3D-printed. 20 images of the same soap film were shot for image processing. A 'Newton's Ring' from the school laboratory was used as a scaling device. In the data analysis process, the distance of the first ring to the center was measured with a vernier caliper to enable the calculation of the diameter of the sight through the ocular lens. Images were shot with a 12 mega-pixel high resolution camera.

The first step of image processing was to separate interference results of different wavelengths of light. The spectrum of the Hg lamp was obtained with a spectrometer, and their corresponding RGB value in MATLAB were as follows.

Table 1 MATLAB RGB of Hg spectrum

| Spectrum | Wavelength/nm | R | G | B |
|---|---|---|---|---|
| Green | 546.07 | 0 | 255 | 0 |
| Orange | 576.96 | 120 | 255 | 0 |
|  | 579.07 | 210 | 255 | 0 |
| Purple | 435.84 | 0 | 0 | 255 |

Then the intensity of the light source was calculated with the equation:

$$I_0(\lambda) = I_{max}(\lambda)\frac{(1+R)^2}{4R}, R = \left(\frac{n-1}{n+1}\right)^2. \quad (3.1)$$

$I_0(\lambda)$ is the source intensity of light with wavelength $\lambda$, $I_{max}(\lambda)$ the maximum intensity detected on the soap film and $n$ the refractive index of the liquid.

After this, the RGB value of each pixel was separated by vector decomposition, using the above spectrum-RGBs as base vectors. Calculate for each pixel the possible thickness deduced from the interferometric pattern of light of each wavelength and then find the intersection, eliminate the physically impossible, and then the thickness distribution of a soap film could be obtained. The relation between possible thickness and the interferometric pattern is:

$$I(\lambda) = I_0(\lambda)\frac{4Rsin^2\delta}{(1-R)^2 + 4Rsin^2\delta}, \delta = \frac{2\pi nd}{\lambda}. \quad (3.2)$$

The obtained thickness distribution would then be converted into an effective-refractive-index distribution according to Eq. 2.2.1.7. For it would be daunting task solving transcendental equations for each pixel, interpolation was applied using MATHEMATICA to return a smooth curve.

In the end, the random potential distribution was obtained from the effective-refractive-index distribution according to Eq. 2.2.2.5. Autocorrelation function then returns the correlation length and $\sigma_V$ needed for calculation.

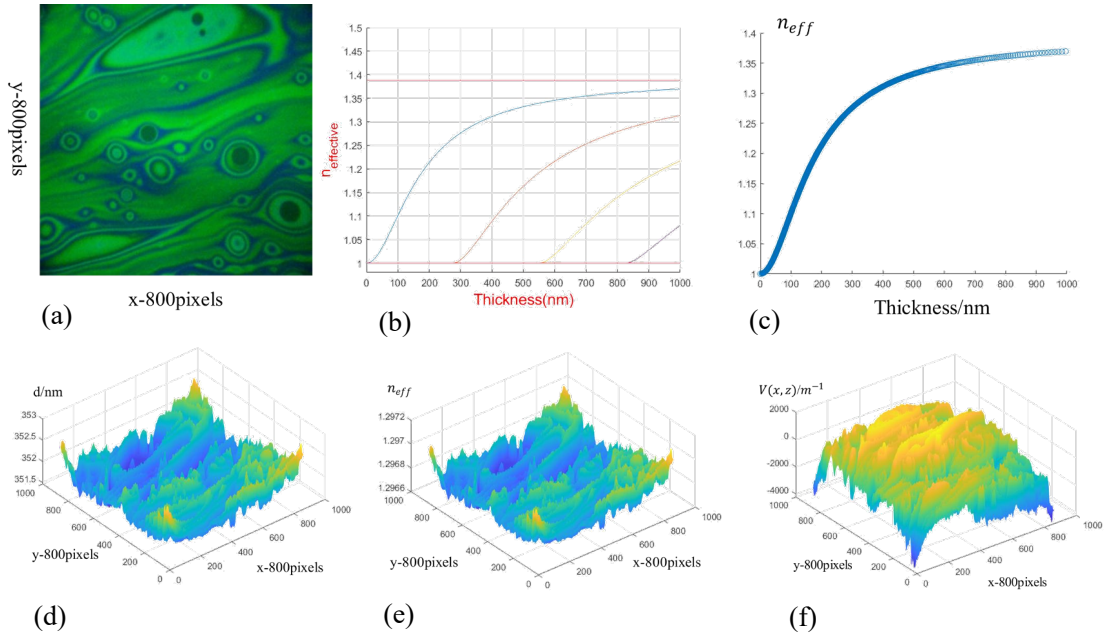

Fig. 8 Data analysis results. (a) the processed image; (b) solving transcendental equation; (c) the first mode only; (d) 3D plot of the thickness distribution; (e) 3D plot of the effective-refractive-index distribution; (f) 3D plot of the random potential distribution

Fourthly, the branched flow pattern was shot in a darkroom environment with 3D-printed holder and a mobile phone with high-speed camera. Statistical data was then obtained using the self-coded .m document, including the decay rate of the mean branch, which is not included in the discussion of this paper. Tracker was used to measure the characteristic length of the branched flow, which in this case was the distance from the coupling point to the first branching.

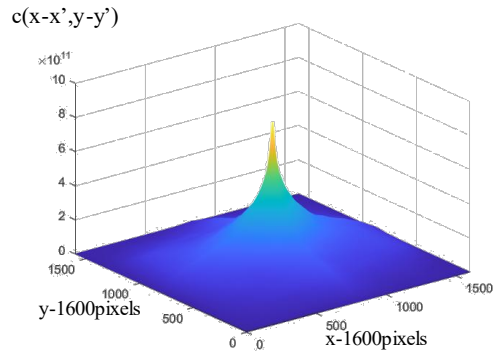

Fig. 9 Autocorrelation of the potential field

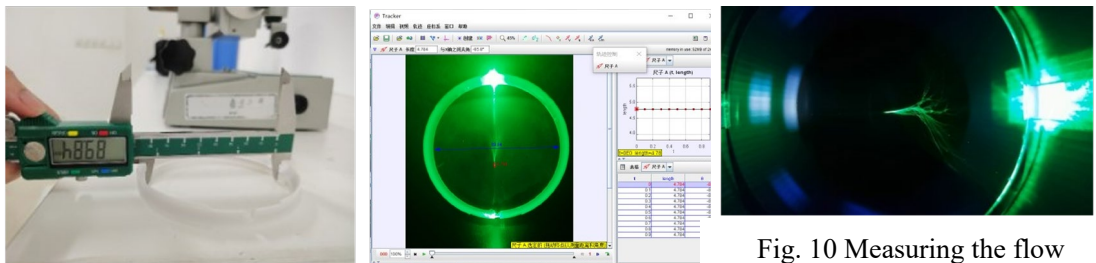

Fig. 10 Measuring the flow

## §4. Experimental Results

Experiments yield that:

$$\bar{n} = 1.2968 \tag{4.1}$$
$$m = k_0 \bar{n} \approx 1.53 \times 10^7 m^{-1} \tag{4.2}$$
$$\sigma_V \approx 9.11 \times 10^5 \tag{4.3}$$
$$l_c \approx 0.202 mm \tag{4.4}$$
$$z = \bar{L} = 5.15 mm \tag{4.5}$$

$$l_c \left(\frac{m}{\sigma_V}\right)^{\frac{2}{3}} \approx 1.33 mm \tag{4.6}$$

The scaling law was proved. This showed that the branching behavior of light from the laser beam was indeed an example of 'branched flow', and that it has the general statistical property—the scaling law—of branched flows.

## §5. Conclusion and Discussion

This paper presents experiments designed on the basis of previous theoretical studies into the topic 'branched flow of light', which used the isomorphism between the governing equations of different systems (wave system and particle system) to transform a wave-problem into a particle-problem and fit into the kick-drift cycle model.

In the experiments, experimental-theoretical analysis, 3D printing/high speed photography/MATLAB image processing & visualization/MATHEMATICA equation solving and other technics were applied to work to prove the scaling law deduced by theoretical study. Although the main process largely follows Ref. [5], for each and every step of realization our study adopted new approaches.

However, there is still room for more. On the one hand, not all the research results are presented in this paper. The left-outs include the heavy-tail distribution of the branch intensity, the scintillation index and so on. On the other hand, researches on branched flow shall not be restricted to light alone. The discovery of the branched flow phenomenon in light has opened up a new approach toward branched flows, and science's further goal would be to apply the conclusions drawn from this approach to a wider range of systems.

**Acknowledgements**

Thanks to School of Physics, Nanjing University for providing funds and equipment.